\documentclass[prl,twocolumn,aps,showpacs]{revtex4}
\usepackage{graphicx}

\begin{document}
\title{
Discrete breathers in nonlinear magnetic metamaterials.
}
\author{N. Lazarides$\ ^{1,2}$, M. Eleftheriou$\ ^{1,3}$ and
G. P. Tsironis$\ ^{1}$
}
\affiliation{
$\ ^{1}$Department of Physics, University of Crete, 
and
Institute of Electronic Structure and Laser,
Foundation of Research and Technology,  
P. O. Box 2208, 71003 Heraklion,  Greece \\
$\ ^{2}$Department of Electrical Engineering,
Technological Educational Institute of Crete,
P. O. Box 140, Stavromenos, 71500, Heraklion, Crete, Greece \\
$\ ^{3}$Department of Music Technology and Acoustics,
Technological Educational Institute of Crete,
E. Daskalaki, Perivolia,
74100 Rethymno, Crete, Greece
}
\begin{abstract}
Magnetic metamaterials composed of split-ring resonators or $U-$type
elements may exhibit discreteness effects in THz and optical frequencies
due to weak coupling.  We consider a model one-dimensional metamaterial
formed by a discrete array of nonlinear split-ring resonators
with each ring interacting with its nearest neighbours.
On-site nonlinearity and weak coupling among the individual array elements
result in the appearence of discrete breather excitations 
or intrinsic localized modes, 
both in the energy-conserved and the dissipative system.
We analyze discrete single and multibreather  excitations, 
as well as a special breather configuration forming a
magnetization domain wall and investigate their mobility
and the magnetic properties their presence induces in the system. 
\end{abstract}
\maketitle
Artificial non-magnetic materials exhibiting magnetic properties
in the Terahertz and optical frequencies have been  recently
predicted theoretically\cite{zhou,ishikawa}
and demonstrated experimentally\cite{yen-moser,katsarakis,grigorenko-enkrich}.
The key element for most of these magnetic metamaterials (MMs) 
is either the split-ring resonator (SRR) or its $U-$shaped 
modification\cite{sarychev}.
The realization of MMs at such (and possibly higher) frequencies 
will affect substantially THz optics and their applications
in devices of
compact cavities, adaptive lenses, tunable mirrors, isolators and converters.  
Moreover, MMs with negative magnetic response can be combined 
with plasmonic wires
that exhibit negative permittivity, producing left-handed materials (LHM),
i.e. metamaterials with negative magnetic permeability $\mu$ and dielectric 
permitivity $\epsilon$ leading to a negative index of refraction 
\cite{drachev-shalaev-zhou,smith,shelby,parazzoli,pendry}.
In the present work we focus entirely on MMs that
have the additional features of being nonlinear as well as discrete.  
While nonlinearity  results in self-focusing, 
discreteness induces localization and, as a result,
the combination of both leads in
the generation of nonlinearly localized modes of the type
of Discrete Breather (DB)\cite{sievers,mackay,marin,marin1,chen}.  
These modes act like stable impurity modes
that are dynamically generated and may alter propagation and
emission properties of the system.

We consider a planar one-dimensional (1D) array of $N$ identical
SRRs with their axes perpendicular to the plane; each unit is equivalent
to an RLC oscillator with self-inductance $L$, 
Ohmic resistance $R$ and capacitance $C$.
The units become nonlinear due to the Kerr 
dielectric that fills their gap and has permittivity equal to
$\epsilon (|{\bf E}|^2) = \epsilon_0 \left( \epsilon_\ell 
          + \alpha {|{\bf E}|^2}/{E_c^2} \right)$,
where ${\bf E}$ is the electric component of the applied electromagnetic
field,
$E_c$ is a characteristic electric field,
$\epsilon_\ell$ the linear permittivity,
$\epsilon_0$ the permittivity of the vacuum, and $\alpha=+1$ ($\alpha=-1$)
corresponding to self-focusing (-defocusing) nonlinearity
\cite{zharov,obrien,lazarides}.
As a result, the SRRs aquire a field-dependent capacitance
$C ( |{\bf E}|^2 ) = \epsilon ( |{\bf E}_g|^2 ) A / d_g$, 
where $A$ is the area of the cross-section of the SRR wire,
${\bf E}_g$ is the electric field induced along the SRR slit,
and $d_g$ is the size of the slit.
Since  $C(U_n)=dQ_n / dU_n$,
the charge $Q_n$ stored in the capacitor of the $n$-th SRR is
\begin{eqnarray}
  \label{1}
   Q_n = C_\ell \left( 1 + \alpha \frac{U_n^2}{3 \epsilon_\ell \, U_{c}^2} 
     \right) U_n, ~~n=1,2,...,N ,
\end{eqnarray}
where
$C_\ell=\epsilon_0 \epsilon_\ell A/d_g$ 
is the linear capacitance, $U_n =d_g E_{gn}$ 
is the voltage across the slit of the $n$th SRR,
and $U_{c} =d_g E_{c}$.
Neighbouring SRRs are coupled due to magnetic dipole-dipole interaction
through their mutual inductance $M$, which decays as the cube of the distance.
For weak coupling between SRRs in a planar configuration, 
it is a good approximation
to consider only nearest neighboring SRR interactions. 
Then, the dynamics of $Q_n$ and the current $I_n$ circulating in the $n$th SRR
is described by
\begin{eqnarray}
  \label{2}
   \frac{dQ_n}{dt} = I_n \\  
  \label{3}
   L \frac{dI_n}{dt} + R I_n + f (Q_n)=
   M \left(\frac{dI_{n-1}}{dt}+\frac{dI_{n+1}}{dt} \right)
  + {\cal E} , 
\end{eqnarray}
where ${\cal E}$ is the electromotive force (emf) 
induced in each SRR due to the applied field,
and $f (Q_n) = U_n$ is given implicitly from Eq. (\ref{1}).
The value of ${\cal E}$ at a given instant is proportional to the magnetic
field component of the applied field perpendicular to the SRR plane, 
and/or the electric field component parallel to the side of the SRRs
which contains the slit\cite{katsarakis1}.
Using the relations 
$\omega_\ell^{-2} = L  C_\ell$,  
$\tau=t  \omega_\ell$,
$I_{c} = U_{c}  \omega_\ell  C_\ell$, 
$Q_{c}=C_\ell  U_{c}$, 
${\cal E} = U_{c} \varepsilon$,
$I_n=I_{c} i_n$, $Q_n = Q_{c} q_n$,  
Eqs. (\ref{2}) and (\ref{3}) can be normalized to 
\begin{eqnarray}
  \label{7}
  \frac{d q_n}{d\tau} &=& {i_n} , \\
  \label{8}
  \frac{d}{d\tau} \left( \lambda \, i_{n-1} - i_n + \lambda \, i_{n+1}
                \right) &=&
		\gamma \, i_n -\varepsilon(\tau) + f (q_n) , 
\end{eqnarray}
where 
$\gamma = R C_\ell  \omega_\ell$ is the loss coefficient, and $\lambda= M/L$
is the coupling parameter. 
In the following,  we use periodic boundary conditions 
(i.e., $i_{N+1}=i_{1}$, $i_{0}=i_{N}$) except otherwise stated.
Analytical solution of Eq. (\ref{1}) for $u_n = f (q_n)$ with the 
conditions of $u_n$ being real and $u_n (q_n=0)=0$, 
gives the approximate expression  
\begin{eqnarray}
  \label{10}
   f (q_n) \simeq q_n -  \frac{\alpha}{3\, \epsilon_\ell} q_n^3 
    + 3 \left( \frac{\alpha}{3\, \epsilon_\ell}\right)^2  q_n^5 
    + {\cal O} (q_n^7)
\end{eqnarray}
That is, the on-site potential $V(q_n)= \int_0^{q_n} f(q_n') \, dq_n'$
is soft for focusing nonlinearity
and hard for defocusing nonlinearity.
Substituting $q_n = A \cos{ ( k D n - \Omega \tau) }$ into the linearized
Eqs. (\ref{7}) and (\ref{8}) with $\varepsilon=0$ and $\gamma=0$,
we obtain the frequency spectrum of linear excitations
\begin{eqnarray}
  \label{14}
    \Omega_k = {[ 1 -2\, \lambda \, \cos(k\, D) ]}^{-1/2} ,
\end{eqnarray}
where $\Omega =\omega/\omega_\ell$ is the normalized frequency, 
$D$ is the separation of neighbouring SRR centers (unit cell size), 
and $k$ the wavenumber ($-\pi \leq k\, D \leq \pi$).

The parameter $\lambda$ can be calculated numerically
for any SRR geometry, since the magnetic field of the current circulating
the SRR is well known. 
Here we estimate $\lambda$ with a simple model\cite{zhou},
neglecting the effects of nonlinearity and coupling on the resonance frequency
\cite{obrien}.
For not very small array dimensions,  
the inductance of a circular SRR of radius $a$ with circular cross-section
of diameter $h$, is $L = \mu_0 a [\ln(16a/h)-1.5]$,
where $\mu_0$ is the permeability of the vacuum.
For a squared SRR with square cross-section with 
side length $\ell=5~\mu m$, 
$t=w=d_g=1~\mu m$ the SRR depth, width, and slit size, respectively,
length of unit cell $D=7 ~\mu m$\cite{katsarakis},
and using that $\ell' = 4(\ell-w)-d_g$ is the length of the axis 
of the wire, $a=\ell'/2\pi$, $h=\sqrt{4 \, w \, t/\pi}$, 
we arrive at $L \simeq 6 \times 10^{-12} ~ H$. 
For this $L$, we evaluate the capacitance necessary for providing
the resonance frequency for a single SRR, 
$f_r \simeq 1/2\pi \sqrt{L C_\ell} = 6.2 ~THz$, 
consistent with the available experimental information\cite{katsarakis},
to be $C_\ell \simeq 11 \times 10^{-17} ~F$.
Consider two neighbouring SRRs (1 and 2) in an array of circular SRRs of 
radius $a$ with circular cross-section of diameter $h$.
The flux $\Phi_2$ threading SRR 2 due to the induced magnetic field in 
SRR 1 $B_1 (r) \simeq \mu_0 S  I_1 / 4  \pi  r^3 +{\cal O}((a/r)^3)$, 
where $I_1$ is the induced current in SRR 1, 
$S=\pi a^2$ is the SRR area and $r$ is the distance from its center ($r \sim D$),
is approximately $\Phi_2 \simeq B_1 (r=D) S$. 
Then, $M = \Phi_2 / I_1 \simeq  \mu_0 S^2 / 4 \pi D^3$
and $\lambda \simeq (\pi/4) (a/D)^3 / [ ln(16a/h)-1.5]$.
For an array of  squared SRRs with square cross-section with dimensions as 
in \cite{katsarakis} we obtain 
$\lambda \simeq (\ell'/D)^3 / 32  \pi^2  [ ln(4\ell'/\sqrt{\pi  w t})-1.5] \simeq 0.02$.
For silver made SRRs, whose conductivity and skin depth are 
$\sigma \simeq 6.15 \times 10^7 ~S/m$ and $\delta \sim 20 ~nm$,
respectively, we obtain 
$R= 2a/\sigma h \delta = \ell'/2\sigma \delta \sqrt{\pi w t} \simeq  3.44$,
and $\gamma \simeq 0.01$.
\begin{figure}[h]
\includegraphics[angle=0, width=.8\linewidth]{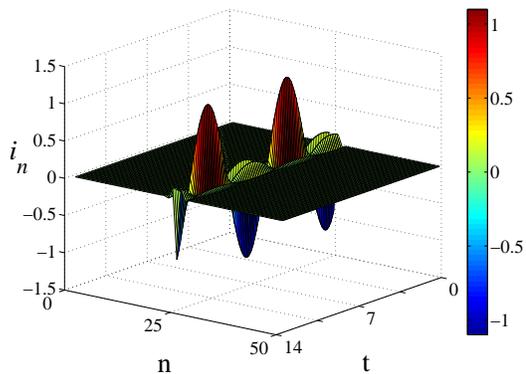}
\caption{
 (color online). Time evolution of a Hamiltonian breather for approximately two periods
 for $\lambda=0.02$,  $T_b=6.69$, $\alpha=+1$, $\epsilon_\ell=2$ and $N=50$.
}
\end{figure}
\begin{figure}[h]
\includegraphics[angle=0, width=80mm, height=35mm]{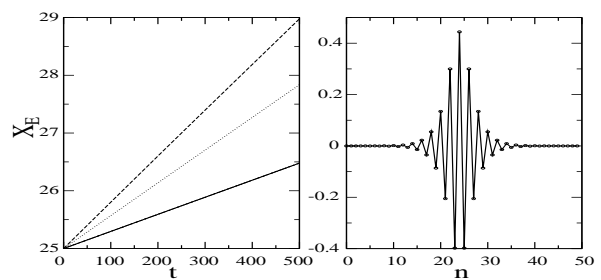}
 \caption{
 Moving Hamiltonian breathers. Right: breather amplitude for 
 $T_b=6.69$, $\alpha=+1$, $\epsilon_\ell=2$, $N=50$, and $\lambda=0.062$
 Left: space-time evolution of the center of energy $X_E$ for the 
 breather shown in the right figure, 
 for $\lambda_p = 0.1$ (solid line);
 $0.2$ (dotted line); $0.3$ (dashed line).
}
\end{figure}

We consider first the lossless case without applied field 
($\gamma=0$, $\varepsilon=0$).
Then, Eqs. (\ref{7}) and (\ref{8}) can be derived from the Hamiltonian
\begin{eqnarray}
  \label{11}
    {\cal H} = \sum_n \left\{  
       \frac{1}{2} \dot{q}_{n}^2  + V ( q_n ) 
        -  \lambda \, \dot{q}_{n}\, \dot{q}_{n+1}
       \right\}  .
\end{eqnarray}    
For Hamiltonian systems DBs may be constructed from the anti-continuous 
(AC)  limit\cite{marin}, where all oscillators are uncoupled
($\lambda \rightarrow 0$), obeying identical dynamical equations.
Fixing the amplitude of one of them (say the one located at $n=n_b$)
to a specific value $q_{b}$, with the corresponding current $i_{b}=0$,
we can determine the oscillation period $T_b$.
An initial condition with $q_{n}=0$ for any $n\neq n_b$, $q_{n_b}=q_b$,
and $\dot{q}_{n}=i_n=0$ for any $n$, represents a trivial DB.
Continuation of this solution for $\lambda\neq 0$ using the Newton's method
\cite{marin}, 
results in numerically exact DBs up to $\lambda_{max}$, 
where the linear excitation frequency band 
(which  expands with increasing $\lambda$)
reaches the DB frequency $\omega_b=2\pi/T_b$.
The linear stability of Hamiltonian DBs is addressed 
through the eigenvalues of the monodromy matrix (Floquet coefficients).
Fig. 1 shows the time evolution of a typical, linearly stable, 
highly localized, DB excitation
($\lambda_{max} \sim 0.067$ for the chosen parameters).
In this figure, plotted vs. time $t$ and array site $n$, is the normalized 
current $i_n$ circulating the $n$th SRR.
Another trivial DB can be obtained for $q_{n}=q_b$ 
for any $n\neq n_b$, $q_{n_b}=0$ and $\dot{q}_{n}=i_n=0$ for any $n$,
corresponding to what we could call a "dark" DB in analogy with the dark soliton
in nonlinear continuous systems.
Such a DB can be continued up to $\lambda \sim 0.025$ but it is linearly 
unstable except for very small $\lambda$.
In order to investigate the mobility of these DBs we followed the procedure 
described  be Chen {\em et al}\cite{chen}.
According to this work, in order to generate a (steady state) moving DB,
having obtained a static DB  $({\bf q}^0, {\bf i}^0={\bf 0})$
by Newton's method,
we integrate Eqs. (\ref{7}) and (\ref{8}) using as initial condition 
$({\bf q}(\tau=0), {\bf i}(\tau=0))
     = ({\bf q}^0, {\bf i}^0={\bf 0}) 
       + \lambda_p  ({\bf 0}, {\bf \delta i})$,
where $\lambda_p$ is the perturbation strength, 
and the perturbation vector ${\bf \delta i}$ corresponds to the current part
of the (normalized to unity) pinning mode eigenvector.
The resulting DB motion is followed by plotting the instantaneous center of 
localization of energy $X_E$ of the DB for several values of $\lambda_p$
and a $\lambda$ value close to $\lambda_{max}$ (Fig. 2);
the parameter $X_E$ is defined as 
\begin{eqnarray}
  \label{11.2}
   X_E = \sum_{n=1}^N n \cdot E_n / E_{tot} ,
\end{eqnarray}
where $E_n$ is the energy at site $n$ and $E_{tot}=\sum_{n=1}^N E_n$. 
We note that Hamiltonian DBs move slowly through the lattice as a 
result of the perturbation. 
Their velocity decreases with increasing $\lambda$, although not uniformly
with $\lambda_p$; in particular, the DB
velocity as a function of $\lambda$ decreases faster as $\lambda_p$ increases. 

In order to generate DBs for the forced and damped system
we start by solving Eqs. (\ref{7}) and (\ref{8}) in the AC limit\cite{marin} 
with emf 
$\varepsilon (\tau) = \varepsilon_0  \sin(\Omega \tau)$.
We identify two different amplitude attractors of the single SRR oscillator,
with amplitudes $q_h \simeq 1.6086$  and $q_\ell \simeq 0.2866$ 
for the high and low amplitude attractor, respectively.
Subsequently, we fix the amplitude of one of the oscillators 
(say the one at $n=n_b$) to  $q_h$ and all the others to $q_\ell$
($i_n$ are all set to zero).
Using this configuration as initial condition, we turn on adiabatically
the coupling $\lambda$. 
\begin{figure}[h]
\includegraphics[angle=0, width=.8\linewidth]{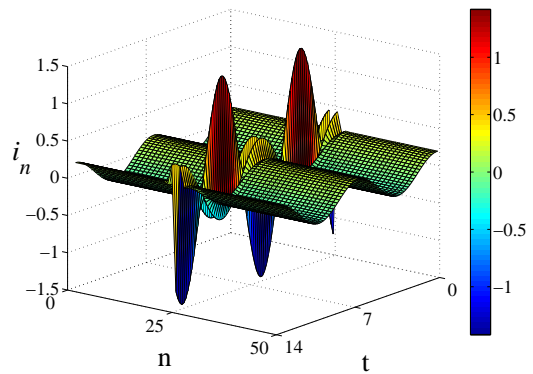}
  \caption{
 (color online).
 Time evolution of a one-site dissipative breather during approximatelly 
 two periods 
 for $T_b=6.82$, $\lambda=0.02$, $\gamma=0.01$, $\varepsilon_0=0.04$,
 $\alpha=+1$, $\epsilon_\ell=2$ and $N=50$.
}
\end{figure}
\begin{figure}[h]
\includegraphics[angle=0, width=.8\linewidth]{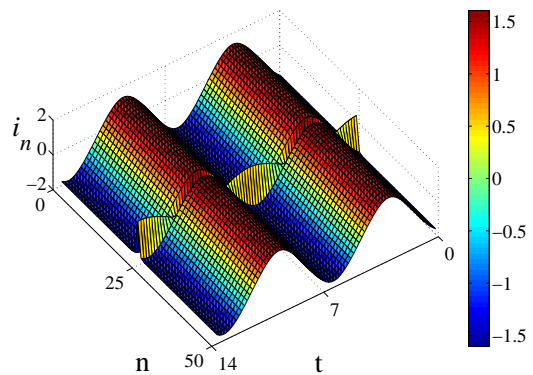}
\caption{
 (color online).
 Time evolution of a one-site dissipative breather of the second type 
 (see text) during approximatelly two periods, 
 for $\lambda=0.01$ and the other parameters as in Fig. 3.
}
\end{figure}
\begin{figure}[h]
\includegraphics[angle=0, width=.8\linewidth]{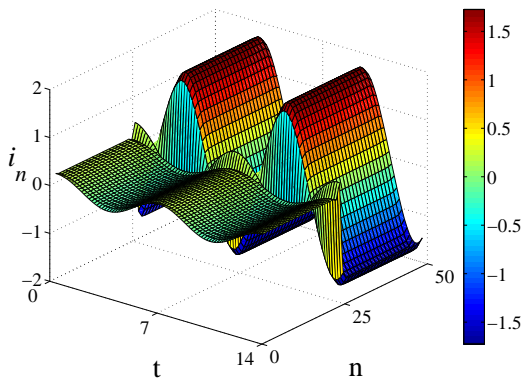}
\caption{
 (color online).
 Time evolution of a domain-wall breather during approximatelly
 two periods.
 Parameters as in Fig. 3.
}
\end{figure}
The initial condition can be
continued for $\lambda \neq 0$ leading to dissipative DB formation
\cite{marin}.
The time evolution of a typical dissipative DB is shown in Fig. 3.
Both the DB and the background are oscillating with different 
amplitudes (high and low, respectively). This should be compared to the 
Hamiltonian DB in Fig. 1, where the background is always zero.
By interchanging $q_h$ and $q_\ell$ in the initial conditions, 
we obtain another DB oscillating with low amplitude,
while the background oscillates with high amplitude (Fig. 4).
With appropriate initial conditions we can also obtain  multi-breathers 
where two or more sites oscillate with high (low) amplitude, 
while the other ones with low (high) amplitude.
Next, we fix the amplitude of half of the SRRs in the array
(say those for $n > N/2$) to $q_h$  and the others to $q_\ell$, 
and  integrate  Eqs. (\ref{7}) and (\ref{8}) from the AC limit with open-ended 
boundary conditions (i.e., $i_{N+1}=i_0=0$). 
In this way we obtain an oscillating domain-wall, as shown in Fig. 5.
\begin{figure}[h]
\includegraphics[angle=-90, width=.7\linewidth]{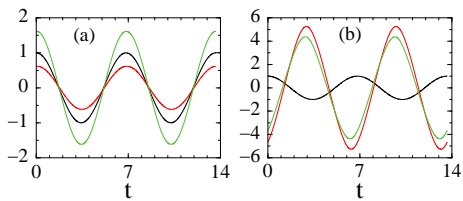}
\caption{
 (color online). Time evolution of $\kappa i_n (\tau)$ (red curve),
 compared with $\cos(\Omega \tau)$ (black curve), and their sum (green curve),
 for two SRRs of the domain-wall DB excitation. 
 (a) low amplitude current oscillation SRR ($n=15$); 
 (b) high  amplitude current oscillation SRR ($n=35$).
 Parameters as in Fig. 3. 
}
\end{figure}

The Hamiltonian DBs investigated are linearly stable, ensuring that 
they are not affected by small amplitude perturbations. 
On the other hand, 
the dissipative DBs are attractors for initial conditions in the 
corresponding basin of attraction and are robust against different kinds of 
small perturbations\cite{marin1}. 
We analysed numerically and confirmed the stability 
of the DBs presented above under various kinds of perturbations; 
we followed the purturbed DB evolution for long time intervals 
(over $2\times 10^4 T_b$)
without observing any significant change in the DB shape\cite{eleftheriou}.

The dissipative system, which includes forcing due to the applied field,
offers the possibility to study its magnetic response.
Assume that the emf is induced by the magnetic component 
$H=H_0 \, \cos(\omega t)$, of the applied field. Then, at least for 
uniform solutions ($I_n=I$), the magnetization $M=S I/D^3$ can be defined.
In the direction perpendicular to the SRRs plane, the general relation 
${\bf B}=\mu_0 ({\bf H} +  {\bf M})$ gives 
\begin{eqnarray}
  \label{12}
      {B}= B_0 (\cos(\Omega \tau) + \kappa i(\tau)) ,
\end{eqnarray}      
where $B_0= \varepsilon_0 U_{c}/ S \Omega \omega_\ell$, and 
$\kappa = \mu_0 S^2 \Omega /  \varepsilon_0 D^3 L$.  
For the material parameters used above  $\kappa \simeq 3$. 
From Eq. (\ref{12}) negative magnetic response appears
whenever the second term in the parentheses is larger in magnitude than
the first one, and has opposite sign. 
Then, one may assign a negative $\mu$ to the medium.
Without nonlinearity, the unique, uniform solution for the SRR array gives 
positive response below the resonance frequency ($\sim\omega_\ell$).
Nonlinearity allows the existence of multiple stable states,
which makes it possible to obtain either positive or negative $\mu$ below 
$\omega_\ell$, depending on the state of the system.
Moreover, exploiting DB excitations, MMs with domains of opposite 
sign magnetic responses can be created.
In Fig. 6 we show the time evolution of $\cos(\Omega \tau)$ and 
$\kappa i_n(\tau)$ as well as their sum, for two SRRs of the domain-wall 
DB (Fig. 5), relatively far from the domain-wall and the ends.
The SRR with low amplitude current oscillation ($n=15$) shows positive 
(paramagnetic) response.
In contrast, the SRR with high amplitude current oscillation ($n=35$)
shows extreme diamagnetic (negative) response, since
$\kappa i_n (\tau)$ is almost out of phase with $\cos(\Omega \tau)$
and much larger in magnitude than that.
Thus, for large enough SRR arrays, one may obtain domain-wall DBs connecting
domains of the array with positive and negative $\mu$.

In conclusion, a 1D planar array of nonlinear SRRs coupled through
nearest-neighbor mutual inductancies was investigated numerically.  
The existence of DBs of various
types, for both the energy-conserved and the dissipative system, was demonstrated. 
We found that longer range interaction does not affect the DB properties 
substantially\cite{eleftheriou}.
We found that Hamiltonian  DBs may be set into uniform motion under a small perturbation.
We also obtained a special DB solution (magnetization domain-wall),
which separates domains of the array with different magnetization. 
Multiple magnetization states are possible in this system due to nonlinearity,
which allows either for negative or positive $\mu$ below resonance.
Moreover, one can exploit multibreathers and domain-wall DBs to create 
MMs with domains of opposite sign magnetic response.
Discreteness effects may appear in SRR arrays with dimensions close to
those reported in \cite{katsarakis}, even though the field wavelength
is much larger than the array dimensions.

We acknowledge  support from the grant "PYTHAGORAS II" (KA. 2102/TDY 25)
of the Greek Ministry of Education and the European Union.

\end{document}